\documentclass[aps,prd,showpacs,amsmath,amssymb,amsfonts,floatfix,superscriptaddress,twocolumn,lengthcheck]{revtex4-2}
\usepackage{graphicx}
\usepackage{subfigure}
\usepackage{verbatim}
\usepackage{dcolumn}
\usepackage{bm}
\usepackage{epsf}
\usepackage{xcolor}
\usepackage{hyperref}
\usepackage{hhline}
\usepackage{float}
\usepackage{enumerate}
\usepackage{bbm}
\usepackage{lipsum}
\definecolor{mygreen}{rgb}{0.01, 0.31, 0.59}
\definecolor{myblue}{rgb}{0.01, 0.31, 0.59}
\hypersetup{
	colorlinks=true,   
	hypertexnames=false,
	linkcolor=blue,  
	citecolor=blue, 
	urlcolor =blue    
}

\begin{document}
\title{Gamma-ray emission from nova shocks expanding in the red giant wind: Interpretation of the 2021 outburst of the recurrent nova RS Ophiuchi}

\author{Jian-He Zheng}
\affiliation{School of Astronomy and Space Science, Nanjing University, Nanjing 210023, People's Republic of China}
\affiliation{Key laboratory of Modern Astronomy and Astrophysics (Nanjing University), Ministry of Education, Nanjing 210023, People's Republic of China}

\author{Yi-Yun Huang}
\affiliation{School of Astronomy and Space Science, Nanjing University, Nanjing 210023, People's Republic of China}
\affiliation{Key laboratory of Modern Astronomy and Astrophysics (Nanjing University), Ministry of Education, Nanjing 210023, People's Republic of China}

\author{Ze-Lin Zhang}
\affiliation{School of Astronomy and Space Science, Nanjing University, Nanjing 210023, People's Republic of China}
\affiliation{Key laboratory of Modern Astronomy and Astrophysics (Nanjing University), Ministry of Education, Nanjing 210023, People's Republic of China}

\author{Hai-Ming Zhang}
\email{hmzhang@nju.edu.cn}
\affiliation{School of Astronomy and Space Science, Nanjing University, Nanjing 210023, People's Republic of China}
\affiliation{Key laboratory of Modern Astronomy and Astrophysics (Nanjing University), Ministry of Education, Nanjing 210023, People's Republic of China}

\author{Ruo-Yu Liu}
\affiliation{School of Astronomy and Space Science, Nanjing University, Nanjing 210023, People's Republic of China}
\affiliation{Key laboratory of Modern Astronomy and Astrophysics (Nanjing University), Ministry of Education, Nanjing 210023, People's Republic of China}

\author{Xiang-Yu Wang}
\email{xywang@nju.edu.cn}
\affiliation{School of Astronomy and Space Science, Nanjing University, Nanjing 210023, People's Republic of China}
\affiliation{Key laboratory of Modern Astronomy and Astrophysics (Nanjing University), Ministry of Education, Nanjing 210023, People's Republic of China}

\date{\today}

\begin{abstract}
Nova outbursts take place in binary star systems comprising a white dwarf (WD) and either a
low-mass Sun-like star (classical novae) or, {a red giant}.
GeV gamma-ray emission has been detected  from  a dozen of classical novae and from one {novae in symbiotic system} (V407 Cyg) by Fermi-LAT. For classical novae, gamma-ray emission is generally thought to be related to internal shocks  formed as fast outflow  collides with the slow outflow. However, {for V407 Cyg}, the origin of the gamma-ray emission has been debated, as both internal shock and external shock,  resulted from the collision between the nova ejecta and the ambient wind of the giant
companion, were suggested to  explain the gamma-ray data. Recently, bright GeV and TeV gamma-ray emission has been detected from a {nova in symbiotic system}, RS Ophiuchi, during its 2021 outburst, which shows a remarkably smooth power-law  decay in time up to about one month after the outburst. We show that this temporal decay behavior  can be interpreted as arising from an adiabatic external shock expanding in the red giant wind. In this interpretation, the gamma-rays are produced by shock-accelerated protons interacting with the dense wind through the hadronic process. We also derive the scaling relations for the decay slopes for both adiabatic and radiative nova shocks in the self-similar deceleration phase. 
\end{abstract}


\maketitle


\section{Introduction}
\label{sec: introduction}
Nova outbursts result from thermonuclear explosions in outer layers of white dwarfs, as they accrete matter from their companions. Ejected material produces shocks, {which occur either as fast components of the nova ejecta  collide with the slower one} (internal shocks) in classical novae, or as the ejecta crashes into a pre-existing  dense wind of the red giant companion  in the  case of {novae in symbiotic system} (external shocks)\cite{Metzger2015,Chomiuk2014,Aydi2020}. 
The presence of shocks in nova outbursts has long been implied by X-ray observations
of hot ($10^7$-$10^8$\,K) presumably shock-heated gas~\cite{O'Brien1994}, observed
weeks to months after eruption~\cite{Mukai2008}. The continuum gamma-ray (0.1-10 GeV)
emission observed by Fermi-LAT from over a dozen Galactic novae provides clear evidence for the acceleration of relativistic particles
by shocks ~\cite{Abdo2010,Ackermann2014,Cheung2016b,Li2017b,Aydi2020b}. 
{The first gamma-ray detection was from nova V407 Cyg in the symbiotic system in 2010 }~\cite{Abdo2010}, where the shocks were
interpreted as the collision between the nova ejecta and the dense wind of the Mira giant
companion (e.g., Refs.~\cite{Abdo2010,Martin2013,Nelson2012} ). However, essentially all of the other high-confidence gamma-ray detections have been classical novae with main sequence companions. In a classical nova, the winds from main sequence stars are weak, and high-energy particle acceleration is suggested to be related
to internal shocks formed in the inhomogeneous ejecta itself (e.g., Ref.~\cite{Metzger2014,Metzger2015,Martin2018}; for a review, see Ref.~\cite{Chomiuk2020}). In fact, Martin et al. 
suggest that internal shocks could even give rise to the majority of the gamma-ray
emission from V407 Cyg, without needing to invoke any interaction with the giant wind~\cite{Martin2018}.

RS Ophiuchi (RS Oph) is a recurrent {nova in a symbiotic system} which displays major outbursts every
15-20 years~\cite{Dobrzycka1994}. The latest outburst, in August 2021, was promptly reported in optical~\cite{Geary2021} and high-energy (0.1-10 GeV) gamma-rays by
Fermi-LAT~\cite{Cheung2021}. Following these alerts, MAGIC and H.E.S.S. observed RS Oph
and detected very-high-energy (VHE)  gamma-ray emission~\cite{MAGIC2022,HESS2022}. The MAGIC observations reveal VHE emission contemporaneous to the Fermi-LAT, and a decrease below
the VHE detection limit two weeks later. H.E.S.S. detect VHE gamma-rays  up to a month after
the  outburst. The H.E.S.S. Collaboration report that, after the peak, the decay 
of the gamma-ray emission with time was well fit with a power-law $t^{-\alpha_{\rm FIT}}$ with  slopes of $\alpha_{\rm H.E.S.S.} = 1.43\pm0.18$ and $\alpha_{\rm LAT} =1.31\pm0.07$, respectively, for H.E.S.S. and LAT observations~\cite{HESS2022,HESSfirst}.
The remarkably smooth power-law decay  is not seen in previous nova outbursts \cite{Ackermann2014,Chomiuk2020}. In this paper, we study the temporal behavior of the gamma-ray emission resulted from relativistic protons accelerated by nova shocks expanding in the red giant wind. We find that the power-law decay of the gamma-ray emission from RS Oph can be well interpreted as  arising from an adiabatic external shock, which is expanding  in the  wind of the companion  star.

 The paper is organized as follows. In sec.~\ref{sec: shock dynamics}, we study whether the nova shock is radiative or adiabatic when it is expanding in the red giant wind, and present the expected light curves of gamma-ray emission in the hadronic model for both  radiative and adiabatic shocks. Then, in sec.~\ref{sec: 2021 RSO},  we apply the theory to the case of the 2021 outburst of RS oph and present an interpretation of the light curve and spectra of the gamma-ray emission. Finally, we give a discussion in sec.~\ref{sec:discussion}.

\section{The shock dynamics  and light curves of the gamma-ray emission}
\label{sec: shock dynamics}
\begin{figure}
\centering
  \includegraphics[width=0.50\textwidth]{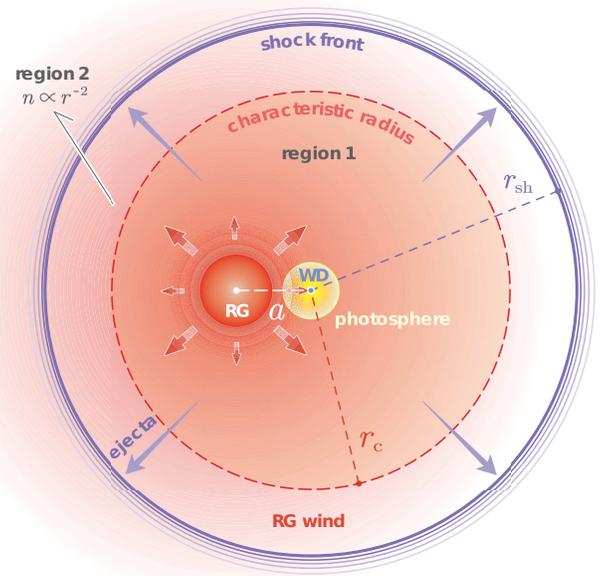}
  \caption{Schematic representation of the expanding shock and the structure of the ambient density around  RS Oph during an outburst. The characteristic radius $r_{\rm c}=3a$ indicates a critical radius that the density profile  changes from  uniform density (region 1) to stellar wind (region 2). }
  \label{fig:RSO}
\end{figure}

The  dynamics of the shock wave resulted from the interaction between a high-velocity nova ejecta and  the surrounding stellar wind has been discussed in~\cite{Bode1985}. It behaves like a
scaled-down supernova remnant~\cite{Chevalier1982}, but with a lifetime of only 
a few weeks or months. The interaction between the ejecta and the wind produces two shocks, a forward shock propagating into the wind and a reverse shock propagating back into the ejecta. Material that has passed
through the reverse shock is quickly cooled and subject to Rayleigh-Taylor instability~\cite{Bode1985}. As we aim to interpret the observed long-lasting gamma-ray emission from RS Oph, we  consider only the gamma-ray emission from the forward shock. The early evolution of the forward shock is characterized by a free-expansion phase,
where the ejecta expands freely and the shock moves at a constant
speed into the wind. The wind density is shaped by  $\rho \propto (r^{2}+a^{2}-2ar \cos\theta)^{-1}$, where $a$ is the semi-major axis of this binary system,$r$ is the radius from the WD (see the Fig.~\ref{fig:RSO}) and $\theta$ is the inclination angle. So, at small radius, the wind density is not spherical centering at the WD position.  Nevertheless, when $r\gg a$,  the density is close to  $\rho\propto r^{-2}$.   The wind density is given by $\rho=A r^{-2}$, where $A=\dot{M}/(4\pi v_{\rm w})=5\times 10^{12} A_{\star}\,{\rm g\,cm^{-1}}$.  The
reference value of $A_{\star}$ corresponds to a mass loss rate of $\dot{M}=10^{-6}M_{\odot}\,\mathrm{yr}^{-1}$ and a wind velocity of $v_{\mathrm{w}}=10\,{\rm km\,s^{-1}}$.
The ejecta will be decelerated when the swept-up mass from the wind is comparable to the ejecta mass. For a mass of $M_{\rm ej}=10^{-6}{M_{\odot} }$, this occurs at
$r_{\rm dec}={M_{\rm ej}}/({4\pi A})=3.2\times 10^{13}\, A_{\star}^{-1} \left(\frac{M_{\rm ej}}{10^{-6} M_\odot}\right)\, {\rm cm}$, corresponding to  a time of
\begin{equation}\label{tdec}
\begin{split}
t_{\rm dec}&=\frac{r_{\rm dec}}{v_{\rm sh,0}}\\
&=0.8 \, A_{\star}^{-1} \left(\frac{M_{\rm ej}}{10^{-6} M_\odot}\right) \left(\frac{v_{\rm sh,0}}{4500\,{\rm km\,s^{-1}}}\right)^{-1}\,{\rm day},
\end{split}
\end{equation}
where $v_{\rm sh,0}$ is the initial shock velocity. We take a reference value of $v_{\rm sh,0}=4500\, {\rm km \,s^{-1}}$ since the initial shock velocity of RS Oph is estimated to be $4200$-$4700\,{\rm km \,s^{-1}}$~\cite{MAGIC2022}.

The bulk of the shock energy is transferred
to thermal plasma. The post-shock temperature of the thermal plasma
is given by
\begin{equation}
T_{\rm sh}=\frac{3\mu m_{\rm p} v^2_{\rm sh}}{16k_{\rm B}} =1.7\times10^7\, \left(\frac{v_{\mathrm{sh}}}{10^3\,{\mathrm{km\,s}^{-1}}}\right)^2\,\mathrm{K},
\end{equation}
where $k_{\rm B}$ is the
Boltzmann constant, $\mu=0.76$ is the mean molecular weight appropriate for nova composition~\cite{Schwarz2007}, and $v_{\mathrm{sh}}$ is the shock velocity.
The thermal plasma cools via the free-free emission and line cooling in the temperature range of $10^7$-$10^8\,{\rm K}$. The cooling rate via free-free emission is $\Lambda_{\rm{ff}}=2\times 10^{-27} ({T_{\rm sh}}/{\rm K})^{1/2} {\rm erg\,cm^3\, s^{-1}}$, 
while the cooling rate via the line emission is
$\Lambda_{\rm line}=3\times 10^{-23} ({T_{\rm sh}}/{10^7\,\rm K})^{-0.7} {\rm erg \,cm^3\,s^{-1}}$~\cite{Schure2009}.
The transition temperature for the two cooling regimes is $T_{\rm c}\approx5\times 10^7\,{\rm K}$, corresponding to a shock velocity of $v_{\rm sh,c}\approx1.7\times 10^3\,{\rm km\,s^{-1}}$.

The cooling time of the shock is $t_{\rm cool}=3k_{\mathrm{B}}T_{\rm sh}/(2 n_{\rm sh}\Lambda_{\rm tot})$, where $n_{\rm sh}=4 n_{\rm w}$ is the shock density and $\Lambda_{\rm tot}=\Lambda_{\rm ff}+\Lambda_{\rm line}$ is the total cooling rate.
For a high temperature, the electron cooling is dominated by the free-free emission. In this case, the cooling time of the shock at the deceleration radius is
\begin{equation}
\begin{aligned}
t_{\rm cool,dec}
&=\frac{9m^{2}_{\rm p}v^{2}_{\rm sh,0}M^{2}_{\rm ej}}{2048 \pi^{2} A^{3}\Lambda_{\rm ff}}\\
&=2.2\, M^{2}_{\rm ej,-6}A_{\star}^{-3}\left(\frac{v_{\rm sh,0}}{4500\,{\rm km \,s^{-1}}}\right)\,{\rm day}.
\end{aligned}
\end{equation}
We define the cooling efficiency behind the shock as the ratio between the dynamic time and the cooling time, i.e. $\eta=t/t_{\rm cool}= r_{\rm sh}/(v_{\rm sh} t_{\rm cool})$.  There are two limiting cases, depending
on whether each shock is radiative $\eta\gg 1$ or adiabatic $\eta\ll 1$~\cite{Metzger2016}. The cooling efficiency at the deceleration time is
\begin{equation}
\begin{aligned}
\eta(t=t_{\rm dec})
&=\frac{512\pi A^{2}\Lambda_{\rm ff}}{9m^{2}_{\rm p}v^{3}_{\rm sh,0}M_{\rm ej}} \\
&=0.36 A_{\star}^2 M^{-1}_{\rm ej,-6} \left(\frac{v_{\rm sh,0}}{4500 \,{\rm km\,s^{-1}}}\right)^{-2}.
\end{aligned}
\end{equation}
This indicates that the cooling efficiency of the shock at the deceleration time increases with the density of the red giant wind, while it decreases with the initial kinetic energy of the ejecta $E_{\rm k}=M_{\rm ej} v_{\rm sh,0}^2/2$.

If the initial kinetic energy of the ejecta is low and the wind density is high (i.e., $\eta(t_{\rm dec})\gg1$), the shock is radiative at the deceleration time.
For a radiative shock, we have (see Appendix \ref{S1} for details)
\begin{equation}
v_{\rm sh}\propto t^{-1/2},  r_{\rm sh}\propto t^{1/2}~~({\rm radiative~shock})
\end{equation}
after the deceleration time. Then we have $t_{\rm cool}\propto v_{\rm sh} r_{\rm sh}^2\propto t^{1/2}$ and thus $\eta$ increases with time, if the free-free emission cooling is dominated. As the temperature of the shock drops with time, the cooling mechanism will  transit to the line cooling regime.
During this regime, $t_{\rm cool}\propto v_{\rm sh}^{3.4} r_{\rm sh}^2\propto t^{-0.7}$ and thus $\eta$ increases with time. Since  the cooling efficiency increases with time in both cooling regimes,  the shock keeps to be radiative.

On the other hand, if the initial kinetic energy of the ejecta is high and the wind density is low (i.e., $\eta(t_{\rm dec})\ll1$), the shock is adiabatic at the deceleration time.
For an adiabatic  shock, we have (see Appendix \ref{S1} for details)
\begin{equation}
v_{\rm sh}\propto t^{-1/3},  r_{\rm sh}\propto t^{2/3}~~({\rm adiabatic~shock})
\end{equation}
after the deceleration time. Then we have $t_{\rm cool}\propto v_{\rm sh} r_{\rm sh}^2\propto t$, if the free-free emission cooling is dominated. During this phase, the cooling efficiency $\eta$ is constant, so the shock keeps to be adiabatic. The cooling mechanism of the shock plasma will  transit to the line cooling regime when the temperature drops to $T_{\rm sh,c}$. This occurs at
\begin{equation}
\begin{aligned}
t_{\rm c}&=t_{\rm dec}\left(\frac{v_0}{v_{\rm sh,c}}\right)^3 \\
&=15\,A_{\star}^{-1} M_{\rm ej,-6} \left(\frac{v_{\rm sh,0}}{4500\, {\rm km \,s^{-1}}}\right)^2 \,{\rm day}.
\end{aligned}
\end{equation}
After the cooling mechanism transits to the line emission regime, we have  $t_{\rm cool}\propto v_{\rm sh}^{3.4} r_{\rm sh}^2\propto t^{0.2}$. During this phase, the cooling efficiency increases with time, and thus the shock will become radiative at a late time. This occurs when $t>t_{\rm r}$, with
\begin{equation}
\begin{aligned}
t_{\rm r}
&=t_{\rm c}\left[\frac{1}{\eta(t_{\rm dec})}\right]^{1.25} \\
&=53\, A_{\star}^{-3.5} M^{2.25}_{\rm ej,-6} \left(\frac{v_{\rm sh,0}}{4500\, {\rm km \,s^{-1}}}\right)^{4.5}\,{\rm day}.
\end{aligned}
\end{equation}

{Although leptonic processes are not completely ruled out, the gamma-ray emission from nova outbursts are likely produced by the hadronic process} (e.g., Refs.~\cite{Abdo2010,Metzger2014,MAGIC2022,HESS2022}), where relativistic protons accelerated by nova shocks collide with ambient ions such as protons (i.e., $pp$ interaction), producing pions that decay into gamma-rays. In our case, the ambient material is the dense wind of the red giant.
The luminosity of the gamma-ray emission ($L_{\gamma}$) depends on the shock luminosity ($L_{\rm sh}$), assuming that a constant fraction ($\epsilon_p$) of the shock energy  is used to accelerate relativistic protons, i.e., $L_{\rm p}=\epsilon_{\rm p} L_{\rm sh}$ and
$L_{\rm sh}=(9/8)\pi m_{\rm p} n_{\rm sh} v_{\rm sh}^3 r_{\rm sh}^2$~\cite{Metzger2016}\,(here $L_{\rm p}$ is the luminosity of relativistic protons). The luminosity of the gamma-ray emission also depends on the efficiency of the gamma-ray production in $pp$ interactions. The proton cooling time scale on hadronic $pp$ interactions is $t_{\rm pp}=1/(\sigma_{\rm pp} c\,n_{\rm sh}K_{\rm pp})$, where $\sigma_{\rm pp}=3\times 10^{-26} {\rm cm}^2$ is the
cross-section for proton-proton interaction,  $K_{\rm pp}=0.5$ is the
inelasticity of pion production,  $n_{\rm sh}=4 n_{\rm w}$ is the number density of the shocked wind and $n_{\rm w}$ is the number density of the wind in the upstream. The gamma-ray production efficiency is $f_{\rm pp}=t /t_{\rm pp}\propto t \, r_{\rm sh}^{-2}$ as long as $t_{\rm pp}$ is longer than the dynamic timescale $t$.
For a radiative shock, $L_{\rm sh}\propto t^{-3/2}$ and $f_{\rm pp}\propto t^0$ after the deceleration time, so we have
\begin{equation}
L_\gamma=f_{\rm pp} \epsilon_{\rm p} L_{\rm sh}\propto t^{-3/2}~~({\rm radiative~shock}).
\end{equation}
For an adiabatic  shock, $L_{\rm sh}\propto t^{-1}$ and $f_{\rm pp}\propto t^{-1/3}$ after the deceleration time, so   we have
\begin{equation}
L_\gamma=f_{\rm pp} \epsilon_{\rm p} L_{\rm sh}\propto t^{-4/3}~~({\rm adiabatic~shock}).
\end{equation}

Before the deceleration time, the ejecta expands freely and the shock moves at an almost constant speed. {The luminosity of gamma-rays  will decline as $L_\gamma\propto t^{-1}$ if the number density follows the wind profile $n\propto r^{-2}$. } However, At small radius, the density profile becomes much more complicated. First, the wind density from the red giant becomes anisotropic centering at the WD position. {Second, according to the three-dimension simulations of RS Oph by Walder et al. (Ref.~\cite{Walder2008}, see their Figure 3), the circumstellar density distribution is substantially flatter than  $r^{-2}$  out to several system separations for a large range of inclination angle $\theta$. From the scaling $L_{\gamma}\propto n^2_{\rm sh}r^3_{\rm sh}$, we see that the light curve will rise if the density is flatter than $r^{-1.5}$.} Third, there may be circumstellar density enhancement  due to the accumulation of gas in the orbital plane and
around the WD~\cite{Martin2013}.  The nova shock may become aspherical because of the
anisotropic matter distribution. Thus the light curve of the gamma-ray emission during this phase is more complicated and a sophisticated simulation is needed for a thorough understanding.

\section{Application to the 2021 outburst of  RS Ophiuchi}
\label{sec: 2021 RSO}

A new outburst of the recurrent nova RS Oph occurred on August 08, 2021~\cite{Geary2021}. Fermi-LAT  detected a transient gamma-ray source positionally consistent with the  optical outburst of RS Oph~\cite{Cheung2021}. We  analyzed the Fermi-LAT data and obtained the light curves of RS Oph in the energy range 0.1-100 GeV, as well as the energy spectra in two time intervals when there are simultaneous observations by MAGIC and H.E.S.S. (see Appendix \ref{S3} for details). We compared the light curve data obtained by us with those shown in~\cite{HESS2022} and~\cite{MG2022} in Fig.~\ref{fig:fermilc} of Appendix \ref{S3}. Our results are generally consistent with theirs. {Defining the initial time $T_{0}={\rm MJD}\ 59434$}, we fit the light curve data during the decay phase  with a  power-law function. It is well fit by a power-law decay $\propto t^{-\alpha_{\rm LAT}}$ with $\alpha_{\rm LAT}=1.28\pm0.05$ (see Appendix \ref{S3} for details). {The slope is consistent with results obtained by H.E.S.S. Collaboration and MAGIC Collaboration \cite{HESS2022,MG2022}}.

The HESS peak has been interpreted as arising from an increasing maximum proton energy in time \cite{HESS2022}. 
Interestingly, the time when the expansion velocity  starts to decrease 
coincides roughly with the HESS peak time (see Extended Data Fig. 7 in Ref.\cite{MAGIC2022}), both of which are around 4-5 days after $T_{0}$. Motivated by this, we propose that the HESS peak time  corresponds to the transition from the ejecta-dominated phase to the deceleration phase, at which the maximum energy of shock-accelerated protons may reach a peak. From the theoretical point of view,  the shock radius is increasing while the shock velocity is  constant during the ejecta-dominated phase, so the maximum particle energy may increase with time, given an appropriate form for the magnetic field evolution. After the deceleration time, the shock velocity starts to decrease, so the maximum particle energy could decrease. Nevertheless, we have insufficient knowledge about the magnetic field evolution as well as the particle acceleration mechanism in nova shocks. A more detailed study of the particle acceleration process in nova shocks would be useful, which is beyond the scope of the present work. 

The peak of the GeV emission may represent the transition of the circum-stellar medium  from a roughly constant density to a wind density profile during the ejecta-dominated phase. The three-dimension simulations of RS Oph by Walder et al. (See their Figure 3 in Ref.~\cite{Walder2008}) have shown that the circumstellar density distribution is substantially flatter than the $r^{-2}$  out to several system separations for a large range of inclination angle.  According to our scaling relation, the gamma-ray luminosity will rise in time if the number density is flatter than $r^{-1.5}$  for a constant velocity shock. 


The HESS gamma-ray peak around $t=4\,$days {after $T_{0}$} implies $M_{\mathrm{ej}}/A_{\star}=5\times10^{-6}{M_\odot}$ for the wind  of the red giant, according to Eq.~(\ref{tdec}). A combination of $A_{\star}=0.4$ and $M_{\mathrm{ej}}=2\times 10^{-6} M_\odot$ for RS Oph during the 2021 outburst is  consistent with the peak time.   For these parameter values, {the cooling efficiency is $\eta=0.03$ } at the deceleration time, so the shock is  adiabatic  and transits to a radiative shock years later. The  decay slope  $1.28\,\pm\,0.05$ of the GeV emission observed by Fermi-LAT agrees with the theoretic  value $t^{-4/3}$ for an adiabatic shock expanding in the wind. The updated decay slope  $\alpha_{\rm H.E.S.S.} = 1.43\,\pm\,0.18$ of the VHE emission observed by H.E.S.S. is also  consistent with this value.

For the above parameter values, the wind density is $n_{\rm w}=4.8\times 10^7\,{\rm cm^{-3}}$ at  the deceleration radius $r_{\rm dec}=1.6\times 10^{14}\,{\rm cm}$.
The cooling time of relativistic protons due to $pp$ interaction is 
$t_{\rm pp}=1.1\times10^7\,{\rm s}$.
At the deceleration time, the gamma-ray production efficiency is $f_{\rm pp}=t_{\rm dec}/t_{\rm pp}=0.03$. The total gamma-ray energy emitted during the 2021 outburst of RS Oph is about $E_\gamma=1.8\times10^{42}\,{\rm erg}$ assuming a distance of 2.45 kpc for RS Oph\cite{MAGIC2022}. About half of the gamma-ray energy is released around the deceleration time, so the total energy of relativistic protons  is estimated to be $E_{\rm p}=0.5E_\gamma/f_{\rm pp}=3\times 10^{43}\,{\rm erg}$. At the deceleration time, about half of the kinetic energy of the nova ejecta has been converted to the shock energy, so the energy conversion efficiency from the shock to relativistic protons is $\epsilon_{\rm p}=E_{\rm p}/(0.5E_{\rm k})\simeq 15\%$, where $E_{\rm k}=M_{\rm  ej}v_0^2/2=4\times10^{44}\,{\rm erg}$ is the initial kinetic energy of the ejecta. This is comparable to the efficiency of accelerating protons in supernova remnant shocks (e.g., Ref.~\cite{Morlino2012}).
\begin{figure}
\centering
  \includegraphics[width=0.51\textwidth]{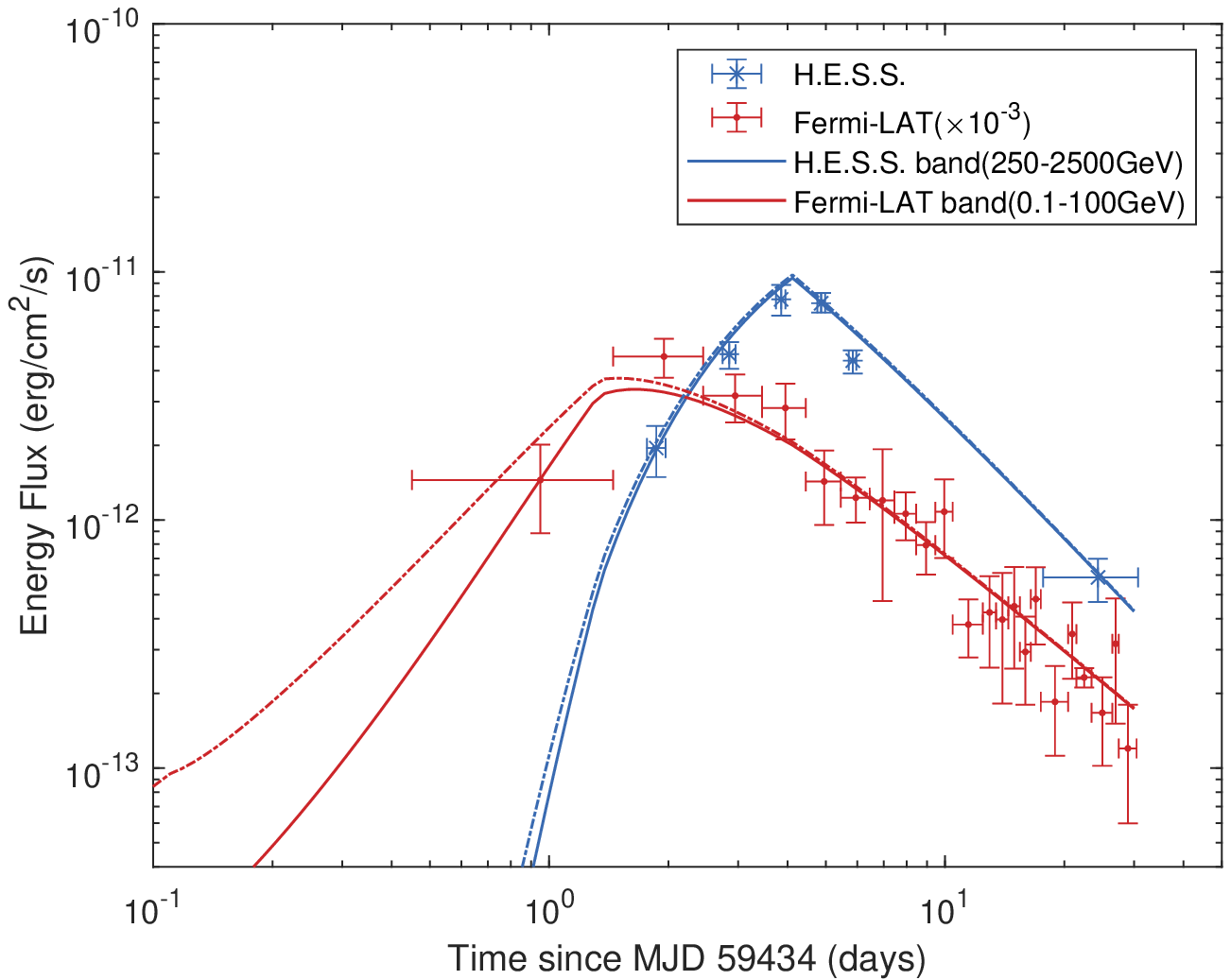}
  \includegraphics[width=0.51\textwidth]{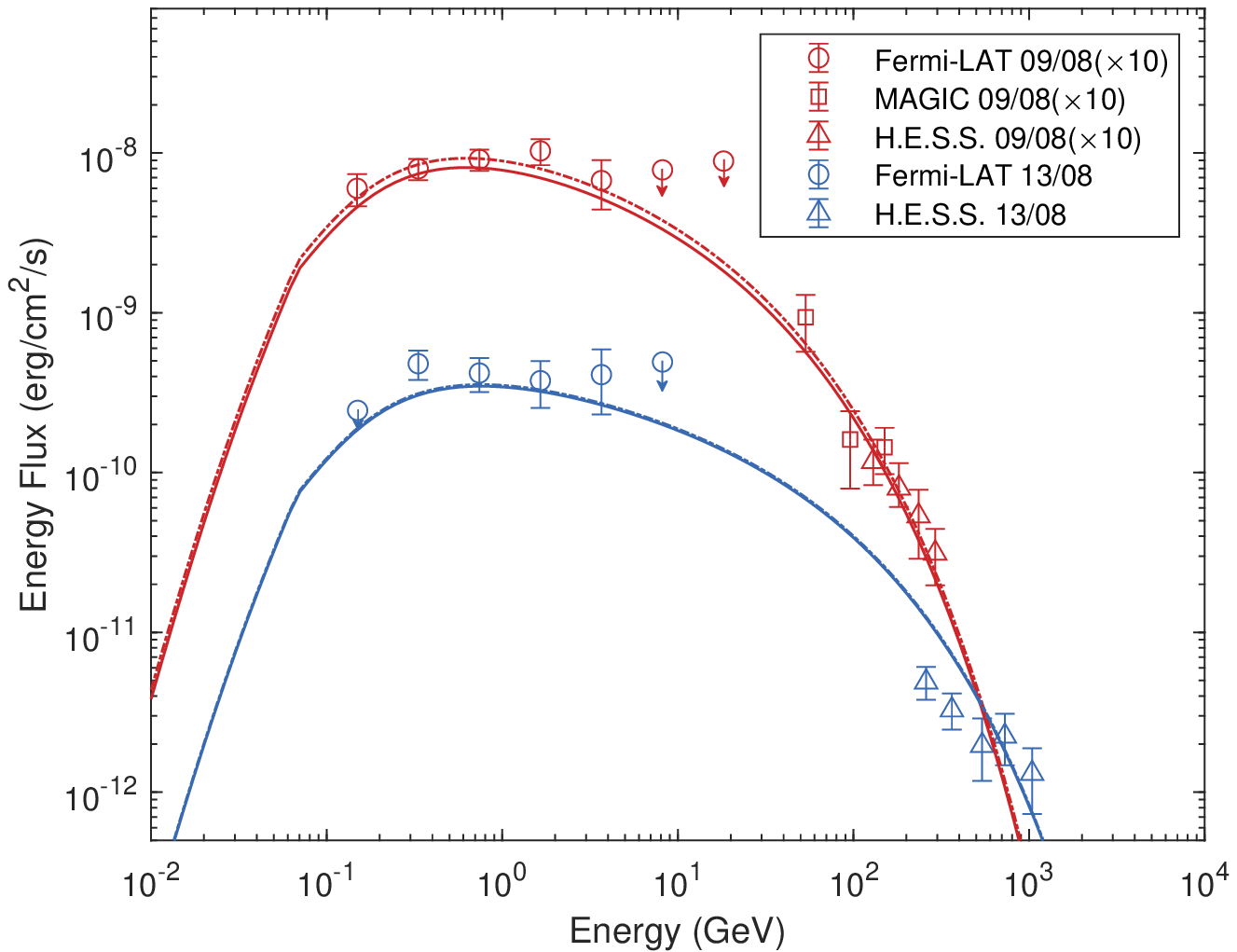}
  \caption{Light curves and energy spectra of the GeV and TeV emission from the 2021 outburst of RS Oph. {\textit{Upper panel}:}  light curves of the gamma-rays emission in Fermi-LAT energy band (0.1-100 GeV) and H.E.S.S. energy band (250-2500 GeV) since $T_{0}=$ MJD 59434. The red  lines represents the model light curves in 0.1-100 GeV and the blue  lines represents the model light curves in 250-2500 GeV. {The solid lines represent the density profile  of $\delta=0$ at small radius and the dash-dotted lines represent the density profile  $\delta=0.5$ at small radius.} {\textit{Lower panel}:} the energy spectra in the two time intervals, August 9, 2021 and August 13, 2021 (corresponding to MJD 59435 and MJD 59439, respectively). The MAGIC and H.E.S.S. data are taken from Ref.~\cite{MAGIC2022} and Ref.~\cite{HESS2022}, respectively.}
  \label{fig:lightcurves}
\end{figure}

The above analytical estimate assumes the self-similar solution for the shock dynamics and treats the gamma-ray production efficiency with a crude approximation. Below we deal with the shock dynamics and hadronic emission process more carefully with a numerical method.
Assuming that the radiation loss of the shock wave is a negligible fraction of the total energy, the
energy conservation gives
\begin{equation}
    E_{\rm tot}=M_{\rm ej}v^{2}_{\rm sh}/2+0.73\,m_{\rm sw}v^{2}_{\rm sh},
\end{equation}
where the first term $M_{\rm ej}v_{\rm sh}^{2}/2$ is the kinetic energy of the ejecta and the second term is the internal energy $E_{\rm sh}$ of the non-relativistic shock driven by the  ejecta~\cite{Blandford1976}. When the ejecta sweeps up the surrounding medium, the initial kinetic energy is transferred into the internal energy of the shocked wind. The mass of the swept-up wind and the radius of the shock evolves with time as
\begin{equation}
\label{equ:dyn}
     \left\{
        \begin{array}{lr}
                \displaystyle \frac{\mathrm{d}m_{\rm sw}}{\mathrm{d}t}=4\pi r_{\rm sh}^{2}\rho (r) v_{\rm sh}\\ \\
        \displaystyle \frac{\mathrm{d}r_{\rm sh}}{\mathrm{d}t}=v_{\rm sh}\\
        \end{array},
  \right.
\end{equation}
We assume the  density profile of the surrounding medium is a broken power-law, motivated by the three-dimension simulations of RS Oph by Walder et al.~\cite{Walder2008}. The density is assumed to be $\rho\propto r^{-\delta}$ at small radii and transits to  $\rho\propto r^{-2}$ at a  critical radius $r_{\rm c}=3a$ (see the Fig.~\ref{fig:RSO}) . { We consider the profiles of $\delta=0$ and $\delta=0.5$ for the density at small radius in our calculation.} 

Protons and electrons are accelerated by the non-relativistic nova shock.
The spectrum of protons accelerated by diffusive shocks can be described by a power-law
\begin{equation}
    \frac{\mathrm{d} N_{\rm p}}{\mathrm{d} E_{\rm p}}=C_{\mathrm{E}}E^{-\alpha_{\rm p}}_{\rm p}{\mathrm{exp} \left[-\left(\frac{E_{\rm p}}{E_{\rm p,max}}\right)^{\beta_{\rm p}}\right]}.
\end{equation}
Assuming a portion ($\epsilon_{\rm p}$) of shock energy was transferred to protons, the normalization factor is given by $\int^{E_{\rm p,max}}_{E_{\rm p,min}} E_{\rm p}({\mathrm{d}N_{\rm p}}/{\mathrm{d}E_{\rm p}})\mathrm{d}E_{\rm p}=\epsilon_{\rm p} E_{\rm sh}$. We take $\alpha_{\rm p}=2.2$, $\beta_{\rm p}=0.5$ and $E_{\rm p,min}=m_{\rm p}c^2$ for the accelerated protons. {\bf Here we take an empirical value of $\beta_{\rm p}=0.5$
simply because it can give a good fit of the spectral data at high energies, without any theoretical
motivation.}   {To explain light curves at VHE, we assume an empirical function for the evolution of the maximum particle energy, i.e.,  $E_{\rm p, max}\propto t^{1.4}$ before $t_{\rm dec}$ and $E_{\rm p,max}\propto t^{-0.2}$ after $t_{\rm dec}$ with a peak value of $E_{\rm p,max}=300 {\rm GeV}$.} The underlying physics for this function is not well-understood, which is due to that we have insufficient knowledge about the magnetic field evolution as well as the particle acceleration mechanism in nova shocks.

The spectrum of gamma-rays generated by pp interactions is calculated following~\cite{Kelner2006},
\begin{equation}
    \frac{\mathrm{d}N_{\gamma}}{\mathrm{d}E_{\gamma}}=c\,n_{\rm sh}\int^{\infty}_{E_{\gamma}}\sigma_{\rm pp}(E_{\rm p})F_{\rm \gamma}\left(\frac{E_{\gamma}}{E_{\rm p}},E_{\rm p}\right)\frac{\mathrm{d}N_{\rm p}}{\mathrm{d}E_{\rm p}}\frac{\mathrm{d}E_{\rm p}}{E_{\rm p}},
\end{equation}
where $F_{\gamma}$ is  the spectrum of the secondary gamma-rays in a single collision. We adopt the delta approximation in calculations.

Based on methods introduced above, we calculate the light curve and the energy spectra of the gamma-ray emission  resulted from $\rm pp$ interactions. The model light curves and energy spectra are shown in Fig.~\ref{fig:lightcurves}.   The parameter values used in the model are $M_{\rm ej}=2\times 10^{-6}M_\odot$, $v_0=4500\,{\rm km \,s^{-1}}$,  $\epsilon_{\rm p}=0.2$ and $A_\star=0.4$~\cite{HESS2022}. These parameter values are consistent with that in the analytical estimate within a factor of two. 
 
The inverse-Compton emission from shock-accelerated electrons is calculated as well (see the Appendix \ref{S3} for details). Assuming the ratio of the energy in relativistic electrons to that in relativistic protons is similar to that for  supernova remnants ($K_{\rm ep} =10^{-4}-10^{-2}$)  (e.g., Ref.~\cite{Morlino2012}) and synchrotron emission from Galactic cosmic rays~\cite{Strong2010}, we find that  the contribution to the gamma-ray flux by shock-accelerated electrons is negligible compared to that contributed by protons.

\section{Conclusions and Discussions}
\label{sec:discussion}
We calculate the light curves of the gamma-ray emission resulted from the nova external shock expanding into the red giant wind. Depending on the kinetic energy of the nova ejecta and the wind density, the nova shock could be radiative or adiabatic at the beginning of the self-similar decelerating phase. An initial adiabatic shock could transit to the radiative shock later on as the temperature of the shock declines with time. In the hadronic model for the gamma-ray emission, we find that the gamma-ray flux decays as a power-law with $t^{-3/2}$   for a radiative shock and  $t^{-4/3}$ for an adiabatic shock  during the self-similar decelerating phase. 
We further find that the gamma-ray emission from the 2021 outburst of RS Oph can be interpreted as arising from an adiabatic external shock which is decelerating in the wind of the red giant. The successful explanation supports the hadronic origin for the gamma-ray emission, indicating  that nova external shocks are able to  accelerate  cosmic ray protons up to TeV energies even when they are significantly decelerated.

{It is  worth mentioning that the possibility of the presence of both external shocks and internal shocks cannot be  ruled out since the decelerated ejecta may be collided by the fast outflow ejected later on. A detailed study of the internal shock contribution is, however, beyond the scope of this work. Interestingly, Gordon et al.  \cite{Gordon2021} investigate 13 gamma-ray emitting novae observed with the Swift Observatory, searching for 1-10 keV X-ray emission concurrent with gamma-ray detections. The discovery that  the only nova in their sample with a concurrent X-ray/gamma-ray detection is also the only  novae in symbiotic systems (V407 Cyg). This exception supports a scenario where novae with giant companions produce shocks with external circumbinary material and are characterized by lower density environments, in comparison with classical novae where shocks occur internal to the dense ejecta that absorbs the concurrent X-ray emission. 
The concurrence of X-ray and gamma-rays is also seen in the 2021 outburst of RS Oph~\cite{Page2022}, favoring the external shock model.}

 \begin{acknowledgments}
This work is supported by the National Key R\&D Program of
China under the Grant No. 2018YFA0404203, the National Natural
Science Foundation of China (grant numbers 12121003, U2031105), China Manned Spaced Project (CMS-CSST-2021-B11). 
 
 \end{acknowledgments}

\begin{appendix}

\section{Shock dynamics in the self-similar deceleration phase}
\label{S1}

For adiabatic shocks, the shock energy is a constant, which gives
\begin{equation}
    m_{\rm sw}v^{2}_{\rm sh}=\frac{4 \pi\rho r^{3-\delta}_{\rm sh}}{3-\delta}\left(\frac{\mathrm{d}r_{\rm sh}}{\mathrm{d}t}\right)^{2}={\rm const}.
\end{equation}
where $\delta$ is the power index of density profile as a function of radius (i.e., $\rho\propto r^{-\delta}$). For a uniform density $\delta=0$, $r_{\rm sh}\propto t^{2/5},v_{\rm sh}\propto t^{-3/5}$;  for the stellar wind case $\delta=2$, $r_{\rm sh}\propto t^{2/3},v_{\rm sh}\propto t^{-1/3}$.

The radiative shock of a nova is similar to  the supernova remnant shock in the snow-plough phase, for which the conservation of momentum gives
\begin{equation}
    m_{\rm sw}v_{\rm sh}=\frac{4 \pi\rho r^{3-\delta}_{\rm sh}}{3-\delta}\frac{\mathrm{d}r_{\rm sh}}{\mathrm{d}t}={\rm const}.
\end{equation}
For a uniform density $\delta=0$, $r_{\rm sh}\propto t^{1/4},v_{\rm sh}\propto t^{-3/4}$; for the  stellar wind case $\delta=2$, $r_{\rm sh}\propto t^{1/2},v_{\rm sh}\propto t^{-1/2}$.

\section{Fermi-LAT data reduction}
\label{S2}
The \textit{Fermi}-LAT is sensitive to $\gamma$-rays with energies from 20 MeV to over 300 GeV, and it has continuously monitored the sky since 2008~\cite{Atwood2009}. The entire sky is monitored every approximately 3 hours by \textit{Fermi}-LAT with a large field of view of about 2.4 sr. 
The  Pass 8 data taken from MJD 59431.45 to 59465.45 are used to study the GeV emission around RS Oph region.
The event class P8R3$\_$SOURCE and event type FRONT + BACK are used.
Only the $\gamma$-ray events in the $0.1-500 \, \rm GeV$ energy range are considered, with the standard data quality selection criteria ``$(DATA\_QUAL > 0) \&\& (LAT\_CONFIG == 1)$".
To minimize the contamination from the Earth limb, the maximum zenith angle is set to be $90^\circ$.
In this work, the publicly available software \textit{Fermitools} (ver. 4.28.5) and \textit{Fermipy} (ver. 1.0.1) are used to perform the data analysis.
Only the data within a $14^\circ\times14^\circ$ region of interest (ROI) centered on the radio coordinates of RS Oph ($R.A. = 267.555^\circ$, $Dec. = -6.7078^\circ$) are considered for the binned maximum likelihood analysis.
The instrument response functions (IRFs) (\textit{$P8R3\_SOURCE\_V3$}) is used.
We include the diffuse Galactic interstellar emission (IEM, $gll\_iem\_v07.fits$),  isotropic emission (``$iso\_P8R3\_SOURCE\_V3\_v1.txt$'' ) and all sources listed in the fourth \textit{Fermi}-LAT catalog~\cite{Abdollahi2020a,Ballet2020} in the background model.
All the normalization and spectral parameters of any sources within $4^\circ$ of the center are set free. The normalization parameters of the IEM and isotropic emission are also set free.

The significance of the GeV emission from the RS Oph is evaluated using the maximum likelihood test statistic (TS), which is defined by $\textrm{TS} = 2 (\ln\mathcal{L}_{1}-\ln\mathcal{L}_{0})$, where $\mathcal{L}_{1}$ and $\mathcal{L}_{0}$ are maximum likelihood values for the background with and without RS Oph (null hypothesis). 
The spectral analysis in the energy range of 0.1-100 GeV for the RS Oph is performed using a binned likelihood analysis. We find that the spectral shape of RS Oph can be reproduced well by a log parabola function with a $\rm TS_{curve}=42.6$ ($\rm TS_{curve}=9$ corresponding to $3\sigma$~\cite{Abdollahi2020a}) during the period from MJD 59435.45 to MJD 59440.45,  where $\rm TS_{curve}$ is defined as $\rm{TS_{curve}}=2(\ln\mathcal{L}_{curved \  spectrum}-\ln\mathcal{L}_{power-law})$.
Hence, we use the log parabola function as the spectral model of RS Oph. The total TS value of RS Oph during the period from MJD 59435.45 to MJD 59440.45 is 2199.3, and the flux in 0.1-100 GeV is $(3.12\pm0.03)\times10^{-6}\rm \ photons\ cm^{-2} \ s^{-1}$. The corresponding one-day-bin SEDs on MJD 59435 and MJD 59439 are shown in the lower panel of Fig.~\ref{fig:lightcurves}.

\begin{figure}
\centering
  \includegraphics[width=0.52\textwidth]{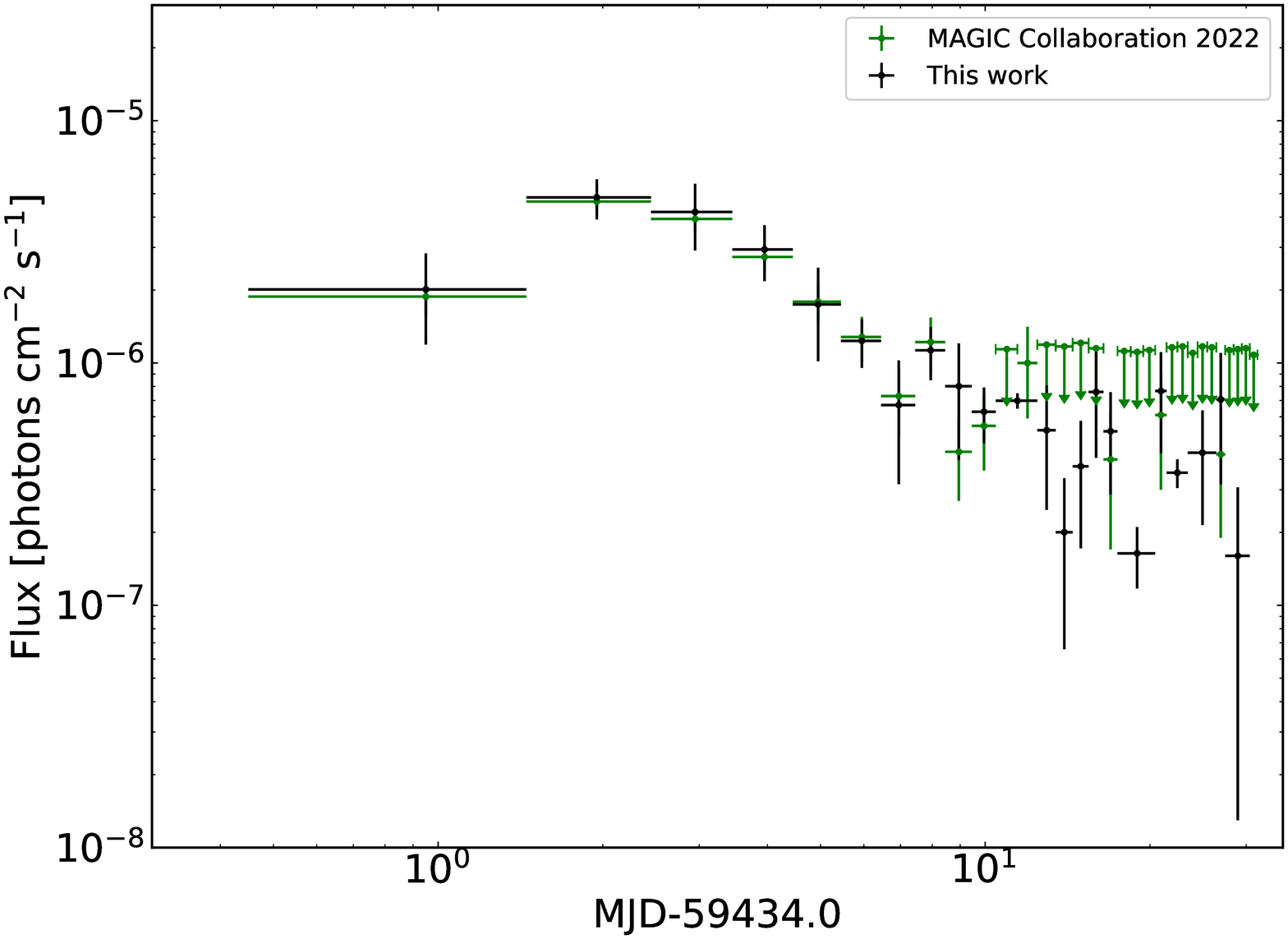}
  \includegraphics[width=0.52\textwidth]{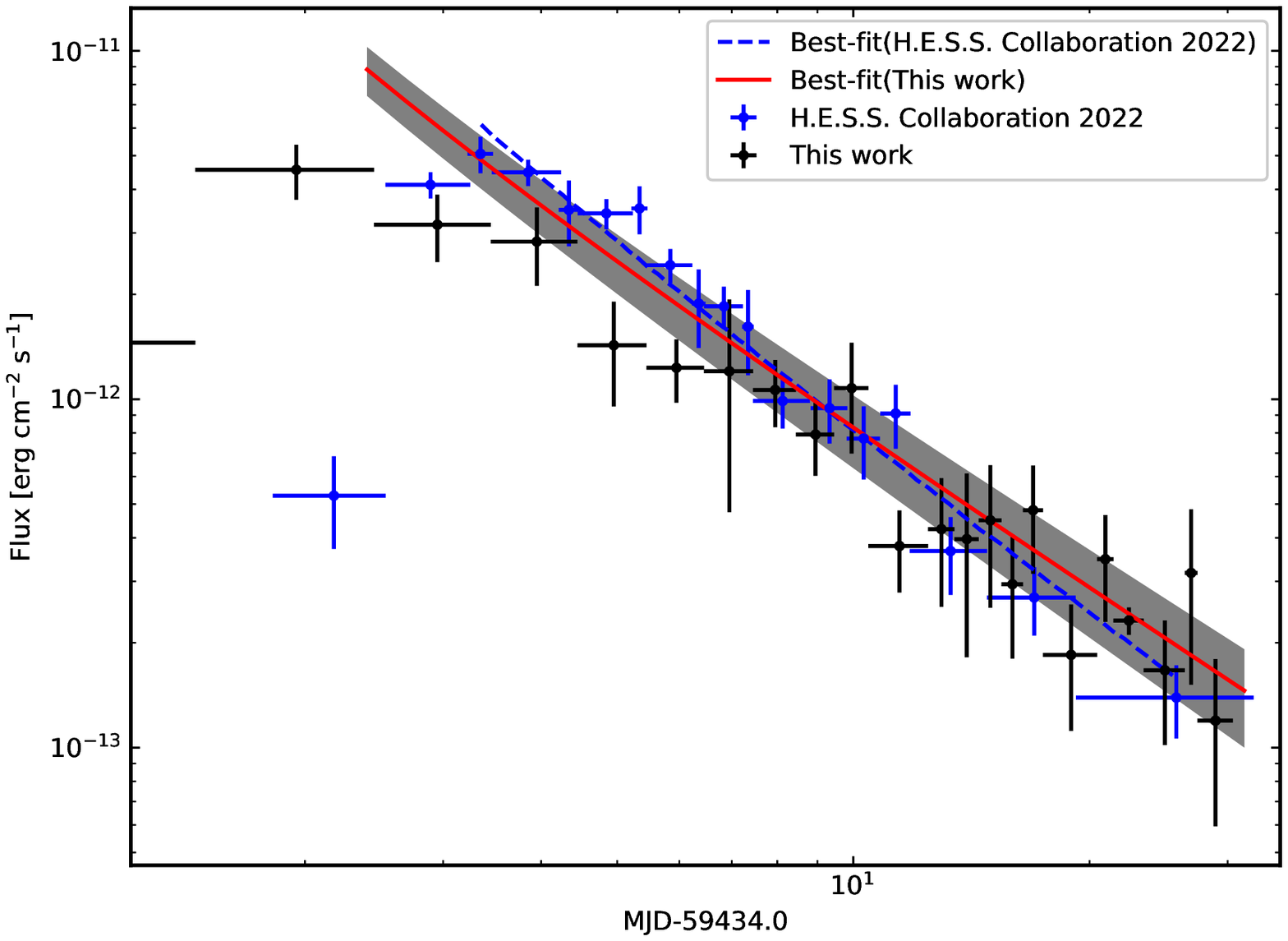}
  \caption{Light curve of GeV emission from RS Oph in 0.1-100 GeV. \textit{Upper panel}: Comparison between our result and the one reported by the MAGIC Collaboration~\cite{MAGIC2022}. \textit{Lower panel}: Comparison between our result and the one reported by the H.E.S.S. Collaboration~\cite{HESS2022}. The red and blue dashed lines represent the best-fit results of our work and the  H.E.S.S. Collaboration~\cite{HESS2022}, respectively. The grey shadow shows $1\sigma$ error range of our best-fit.}
\label{fig:fermilc}
\end{figure}
We generate the light curve of the gamma-ray emission in 0.1-100 GeV by using the adaptive-binning method with the log parabola spectral model. The time bins are generated with constant significance ($\textrm{TS}\geq9$; $\textrm{TS}=9$ approximately corresponds to $\sim 3 \sigma$ detection), and the minimum time-bin is set to be one day. Note that, if the spectral parameter $\beta$ of the time bins is not constrained well, we  fix it to the value ($\beta=0.16$) of the total SED. In the upper panel of Fig.~\ref{fig:fermilc}, we compare our result with the one reported by the MAGIC Collaboration~\cite{MAGIC2022}. Our data agree well with those in the MAGIC Collaboration~\cite{MAGIC2022} before 10 days. After 10 days, the one-day-bin significance drops and the adaptive-binning method becomes effective in our analysis. 
In the lower panel of Fig.~\ref{fig:fermilc}, we compare our light curve with the one reported by  H.E.S.S. Collaboration~\cite{HESS2022}. They are generally consistent with each other, but the best-fit of our light curve data  after the peak gives a slightly shallower decay ($t^{-1.28\pm0.05}$) than that in H.E.S.S. Collaboration~\cite{HESS2022}.
The difference should be attributed to the different choices of the time bin at late time when the significance of the signal decreases. 

\section{Gamma-ray emission from shock-accelerated electrons}
\label{S3}

\begin{figure}
    \centering
    \includegraphics[width=0.51\textwidth]{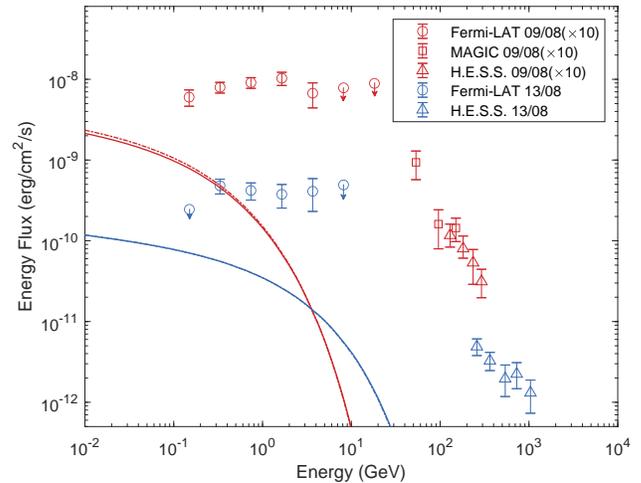}
    \caption{Spectra of the IC emission from shock-accelerated electrons on August 9, 2021 (the red line) and August 13, 2021 (the blue line) assuming $K_{\rm ep}=10^{-2}$. }
    \label{fig:IC}
\end{figure}

Electrons accelerated by shocks can also generate high energy gamma-rays by inverse-Compton (IC) scattering and bremsstrahlung. The cooling time of bremsstrahlung is estimated to be much longer than that of IC emission \cite{MAGIC2022}, so the main cooling channel for electrons is the IC emission. The electron spectrum after cooling is given by
\begin{equation}
    \frac{dN_{\rm e}}{d\gamma_{\rm e}}\propto
     \left\{
        \begin{array}{lr}
                \displaystyle \gamma^{-p}_{\rm e} \qquad \quad \qquad \gamma_{\rm e,m}<\gamma_{\rm e}<\gamma_{\rm e,c}\\ \\
        \displaystyle \gamma^{-p-1}_{\rm e}\exp\left(\frac{-\gamma_{\rm e}}{\gamma_{\rm e,max}}\right) \quad \gamma_{\rm e,c}<\gamma_{\rm e}.
        \end{array}
  \right.
\end{equation}
where $\gamma_{\rm e,m}$ is the Lonentz factor of the minimum energy electrons and $\gamma_{\rm c}=3\pi m_{\rm e}c/(4\sigma_{\rm T}u_{\rm ph}t)$ is the critical Lorentz factor of fast-cooling electrons ($u_{\rm ph}$ is the energy density of soft photons). 
The  electron energy distribution is taken to be ${\rm d}N_{\rm e}/{\rm d}\gamma_{\rm e}=\gamma_{\rm e}^{-2.2}$. The maximum energy of electrons $\gamma_{\rm e,max}$ is obtained by equating the acceleration time with the cooling time, as the cooling time is shorter than the shock dynamic time. This results in a  lower  maximum  energy for electrons than that for protons. In Fig.~\ref{fig:IC}, we plot the spectra of gamma-rays emitted by electrons assuming that the ratio of the energy in relativistic electrons to that in relativistic protons is $K_{\rm ep}=10^{-2}$. In fact,  the ratio is inferred to be $K_{\rm ep}=10^{-4}$-$10^{-2}$ for individual supernova remnants (e.g., Ref.~\cite{Morlino2012}) and synchrotron emission from Galactic cosmic rays~\cite{Strong2010}). For $K_{\rm ep}=10^{-4}$-$10^{-2}$, the gamma-ray flux contributed by electrons is much lower than that contributed by protons.
\end{appendix}


\begin{thebibliography}{59}
\bibitem []{HS2022}
    {If the distance of RS Oph is taken to be 1.4 kpc as in H.E.S.S.~\cite{HESS2022}, the gamma-ray data can also be well explained with small modifications of the parameter values: $M_{\rm ej}=10^{-6}M_\odot$, $A_\star=0.25$.}
\end{thebibliography}
\end{document}